\documentclass[aps,pre,10pt,twocolumn]{revtex4-1}
\usepackage{bm}
\usepackage{graphicx}
\usepackage{amsmath}
\usepackage{amssymb}
\usepackage{amscd}
\usepackage{epstopdf}
\usepackage{verbatim}
\usepackage[usenames]{color}
\usepackage{array}
\usepackage{xspace}
\usepackage{xstring}
\usepackage[normalem]{ulem}
\usepackage{wasysym}
\usepackage{mathrsfs}
\DeclareMathAlphabet{\mathscrbf}{OMS}{mdugm}{b}{n}

\definecolor{Purple}{rgb}{1,0,1}
\definecolor{Red}{rgb}{0.8,0,0}
\definecolor{Green}{rgb}{0,0.5,0}
\definecolor{Blue}{rgb}{0,0,0.8}

\newcommand{\ie}{\textit{i.e.}\xspace}
\newcommand{\eg}{\textit{e.g.}\xspace}
\newcommand{\Dr}{D_\text{r}}
\newcommand{\rw}{\vec{s}}

\newcommand{\R}{\overline{\rho}}
\newcommand{\p}{\overline{p}}

\renewcommand{\vec}[1]{{\bf #1}}
\newcommand{\vecu}[1]{{\hat{\bf #1}}}
\newcommand{\vecg}[1]{{\bm #1}}
\newcommand{\vecug}[1]{{\bf\hat{\bm #1}}}
\newcommand{\tens}[1]{{\bm #1}}

\newcommand{\cov}{\nabla}
\newcommand{\covps}{{\bm\breve{\nabla}}}

\newcounter{Cfp}
\newcommand{\fp}{\ifnum\theCfp=0{Fokker-Planck (FP)}\stepcounter{Cfp}\else{FP}\fi\xspace} %

\newcommand\sect[1]{\vspace{0.4ex minus 0.2ex} {\bf \emph{#1}:}\xspace} 

\newcommand{\SI}{(see appendix)\xspace}

\graphicspath{{figures/}}

\begin{document}

\title{Active Particles on Curved Surfaces}
\author{Yaouen Fily, Aparna Baskaran, Michael F. Hagan}
\affiliation{Martin Fisher School of Physics, Brandeis University, Waltham, MA 02453, USA}

\begin{abstract}
Recent studies have highlighted the sensitivity of active matter to boundaries and their geometries. Here we develop a general theory for the dynamics and statistics of active particles on curved surfaces and illustrate it on two examples. We first show that active particles moving on a surface with no ability to probe its curvature only exhibit steady-state inhomogeneities in the presence of orientational order. We then consider a strongly confined 3D ideal active gas and compute its steady-state density distribution in a box of arbitrary convex shape.
\end{abstract}

\maketitle


Initially introduced as a theory of animal flocking, the paradigm of active materials has grown into a robust framework to describe systems whose building blocks have the ability to move autonomously by extracting energy from their surroundings~\cite{Marchetti2013}. This makes it a natural starting point to understand the mechanics of living systems or engineer energy-consuming biomimetic materials.

One important class of systems in this paradigm consists of active objects moving on a surface~\cite{Sknepnek2015}.
Such cases abound in biology, and the boundaries are rarely flat.
For example, the migration of cells along curved tissues occurs on the gut's lining~\cite{Fatehullah2013,Ritsma2014}, on the surface of an organoid~\cite{Ewald2008}, in a gastrulating embryo~\cite{Vasiev2010}, and in the growing cornea~\cite{Collinson2002}. At the macroscale, animals run and flock along uneven terrains. In the synthetic realm, active microtubule bundles adsorbed at a liquid-liquid interface~\cite{Sanchez2012} or on a lipid membrane~\cite{Keber2014} also exhibit strong curvature effects.

Another important class of systems is confined active materials.
Introducing flat walls suffices to prompt unusual physics~\cite{Ray2014,Mallory2014,Harder2014,Ni2015, Solon2015}, \eg the breakdown of the concept of equation of state~\cite{Solon2015}.
Curved walls lead to even more dramatic behaviors, including spontaneous flow~\cite{Galajda2007,Wan2008,Tailleur2009,Angelani2011,Ghosh2013, Ai2013,Wioland2013,Kantsler2013,Lushi2014,Camley2014,Nikola2015}, concentration of particles in certain regions of space~\cite{Kaiser2012,Kaiser2013,Guidobaldi2014,Fily2014a,Fily2015,Maggi2015,Nikola2015}, and propulsion of passive objects~\cite{Mallory2014a,Kaiser2014}.
In cells, active processes can interact with the membrane's curvature to help the cell change its shape~\cite{Ramaswamy2000,Maitra2014} or locate its own center~\cite{Ma2014}.

Despite the prevalence of curvature in systems of interest, existing
theories of active matter have been limited to specific geometries of
surfaces and confining boundaries. In this work we address this limitation by considering a system of self-propelled particles and i) describing
the generic effects of curvature on the dynamics of such particles at the
level of a Langevin equation and ii) deriving a generic covariant
statistical theory for the dynamics of self-propelled particles that
accounts for curvature induced effects for arbitrary surfaces.
This formalism can be applied to a wide variety of problems, from particles embedded in a curved surface, to particles whose path is temporarily blocked by a wall.
We illustrate this point by studying two cases of interest corresponding to these two situations: a) mutually aligning particles moving on a curved surface and b) a 3D ideal gas of active particles strongly confined in a convex box.
In case a), we show that, at steady state, particles in the isotropic phase have spatially uniform density, independent of curvature. However, upon acquiring orientational order, particles exhibit curvature-driven inhomogeneities in density and polarization.
We compute these inhomogeneities explicitly for weakly curved surfaces, and show that they are nonlocal and quadratic in the height of the surface's deformations.
In case b), we show that strongly confined particles (particles in a container much smaller than the persistence length of their motion) never leave the boundary, that their orientation is nearly aligned with the boundary's normal, and that their density is proportional to the local Gaussian curvature of the boundary, thus extending a previous result obtained in 2D~\cite{Fily2014a,Fily2015}.

\begin{figure}[h]
\centering
\includegraphics[width=0.5\linewidth]{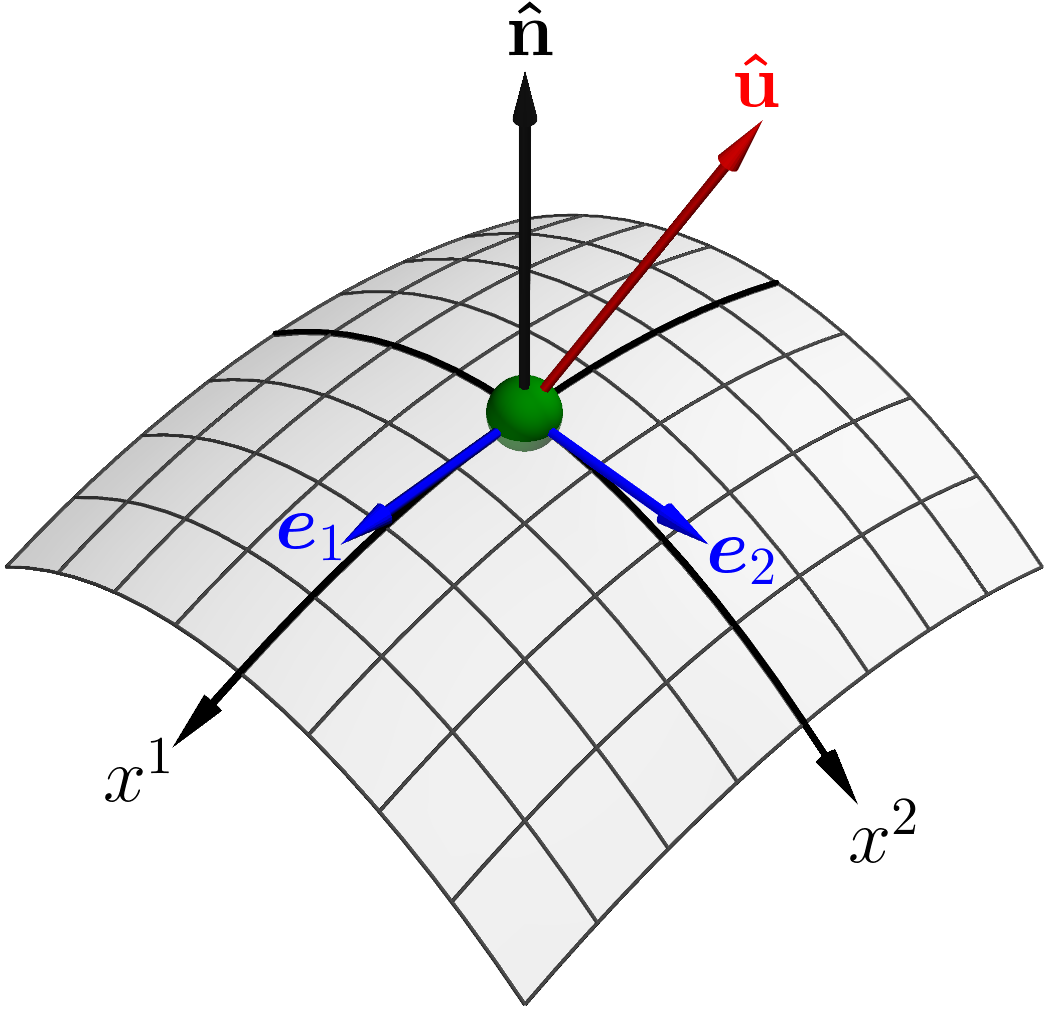}%
\hspace{0.05\linewidth}%
\includegraphics[width=0.37\linewidth]{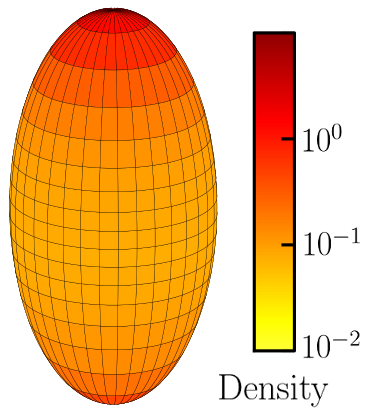}
\caption{ (Color online)
Left:
Geometric notations.
$\vecu{n}$ is the unit normal to the wall.
$\left(\vec{e}_1,\vec{e}_2\right)$ is the local basis of the tangent plane associated with the coordinates $\left(x^1,x^2\right)$.
The particle (green sphere) moves along the projection of its orientation $\vecu{u}$ onto the surface.
Right: Color map of the predicted particle density for a strongly confined 3D nonaligning ideal active gas in a prolate spheroid with aspect ratio 2 (see text).
}
\label{fig:sketch}
\end{figure}


\sect{Statistical Framework}
Let us consider a collection of self-propelled particles on a smooth curved surface parametrized by $\rw(x^1,x^2)$.
Each particle is characterized by its position coordinates $x=\left(x^1,x^2\right)$ and the components $u=\left(u^1,u^2\right) $ of its orientation in the local basis of the tangent plane $\left\{ \vec{e}_a\right\} _{a=1,2}$ where $\vec{e}_a=\partial \rw/\partial x^a=\partial_a\rw$ \
(superscripts/subscripts denote contravariant/covariant indices)~\footnote{For a general purpose introduction to differential geometry in the context of soft matter, see Ref.~\cite{Deserno2015}.}.
The equations of motion have the form
\begin{align}
dx_\alpha^{\ a}/dt=v_\alpha^{\ a}
, \
Du_\alpha^{\ a}/dt=w_\alpha^{\ a}+\sigma_{\alpha\  b}^{\ a}\xi_\alpha^{\ a}
\label{eq:eom}
\end{align}
where Greek indices are particle labels,
$\xi$ is a Gaussian noise with zero mean and correlations $\left\langle \xi_{\alpha}^{\ a}(t) \xi_{\beta b}(t') \right\rangle =2\delta_{\alpha\beta} \delta_{\ b}^{a} \delta(t-t') $ and $v$,
$w$, $\sigma $ are arbitrary functions of the positions and orientations
of the particles. This allows for particles with arbitrary interactions
(forces through $v$ and torques through $w$) and angular noise.
The intrinsic time derivative $\frac{Du^a}{Dt}=du^a/dt+\Gamma^a_{\ bc}(dx^b/dt)u^c$, where $\Gamma^a_{\ bc}=\vec{e}^a\cdot \partial_b\vec{e}_c$ are the Christoffel symbols of the second kind, ensures that the equations of motion are explicitly covariant, \ie, they are by construction independent of the parameterization of the curved surface.

To unfold the influence of curvature on emergent behavior in diverse cases, we seek to develop a statistical description that is also explicitly covariant.
We begin by defining the one-particle scalar probability density
\begin{align}
f(x,u,t) =
\frac{1}{g}
\left\langle \sum_\alpha
\prod_{i=1}^2 \delta\left[ x^a - X_\alpha^{\ a}(t) \right]
\delta\left[ u^a - U_\alpha^{\ a}(t) \right]
\right\rangle
\nonumber 
\end{align}
where $g$ is the determinant of the metric, and the average
is over the particle trajectories $\left\{\left( X_{\alpha }\left( t\right)
,U_{\alpha }\left( t\right) \right)\right\}_\alpha$.
$f$ obeys a \fp equation of the form~\SI
\newcommand{\mf}[1]{\overline{#1}}
\begin{align}
\partial_t f = - \covps_{x^a} \left( \mf{v}^a f \right)
    - \partial_{u^a} \left( \mf{w}^a f \right)
    + \partial_{u^a} \left[ \mf{\sigma}^{ac} \partial_{u^b} \left( \mf{\sigma}_c^{\ b} f \right) \right]
\label{eq:fp}
\end{align}
where $\mf{v}$, $\mf{w}$, $\mf{\sigma}$ are the mean-field versions of $v$, $w$, $\sigma$.
$\covps_{x^a}$ is the phase space covariant spatial derivative, which transforms like a covariant vector, whereas $\partial_{x^a}$ does not.
For a contravariant vector field $z^a$ it reads
\begin{align}
\covps_{x^b} z^a = \partial_{x^b} z^a + \Gamma^a_{\ bc} z^c - \Gamma^d_{\ bc} u^c \partial_{u^d} z^a
\label{eq:covdev}
\end{align}
The first two terms form the usual covariant derivative $\nabla_b z^a$, which accounts for the fact that $z(x,u)$ and $z(x+dx,u)$ live in different tangent planes with different local bases, therefore $z$ can be locally uniform ($\nabla_{x^b}z^a=0$) when its components are not ($\partial_{x^b}z^a\neq0$). Likewise, local basis changes along the surface affect the orientation coordinates $u^a$, which in turn affect $z^a$ as captured by the second $\Gamma$ term. For functions of the position only, the last term vanishes and $\covps$ reduces to $\cov$.

We stress that the covariance we discuss here is different from the one studied by Graham~\cite{Graham1977} but similar to the one discussed by Debbasch and Moreau~\cite{Debbasch2004} in the context of passive diffusion on a curved surface (in which case $u$ is the particle's momentum).
The crucial point is that a choice of position coordinates implies a choice of orientation coordinates through the definition of the local basis $\{\vec{e}_a\}$.
Therefore, we are not interested in covariance under arbitrary changes of the phase space coordinates (as in~\cite{Graham1977}), but rather under arbitrary changes of the position coordinates along with the implied change of orientation coordinates.
Such a situation arises naturally when the phase space of a particle consists of a position on a curved surface and a vector tangent to the surface (orientation, momentum...).

Most importantly, Eq.~\eqref{eq:fp} is a good starting point to derive (covariant) hydrodynamic equations, which are at the very core of the theoretical description of active matter. Hence Eq.~\eqref{eq:fp} is central to understanding active systems on curved surfaces. We now illustrate this point by examining two concrete examples of interest.


\sect{Flocking on a curved surface}
Let us begin with self-propelled particles that are entirely restricted to a curved surface, such as colloids trapped at an interface, or animals on the ground.
In this case the active force is always tangent to the surface, and we can choose $u^a u_a=1$.
We focus on particles that are ``unaware'' of the third dimension, \ie, whose dynamics never refers explicitly to the fact that the surface is embedded in 3D space, and show that curvature still affects their collective behavior.
Even in the absence of noise or interactions, curvature enters the dynamics through the intrinsic time derivative $\frac{Du}{dt}$ in the equation of motion for the orientation [Eqs.~\eqref{eq:eom}]. Similar to free-falling objects in general relativity, such particles follow geodesics of the surface.

Clearly, these deflections have the potential to disrupt the collective dynamics of flocking particles, as illustrated by the band-shaped flocks recently observed on the surface of a sphere~\cite{Sknepnek2015}.
With this in mind, we study a model of mutually aligning self-propelled particles obeying the equations of motion~\eqref{eq:eom} with 
$v_\alpha^{\ a}=v_0 u_\alpha^{\ a}$,
$w_\alpha^{\ a}=\sum_\beta K (x_{\alpha\beta})
    \left( \delta^a_{\ b}-u_\alpha^{\ a}u_{\alpha b}\right) u_{\beta }^{a}$ 
and $\sigma_{\alpha\ b}^{\ a}=\sqrt{\Dr}\left( \delta^a_{\ b} -u_\alpha^{\ a} u_{\alpha b}\right)$.
Here $w$ corresponds to a pairwise aligning torque reminiscent of the XY model of ferromagnetism, and $\sigma$ models the noise from the environment as the projection of an uncorrelated Gaussian white noise onto the curved surface. The kernel $K$ controls the range of the aligning interaction. In the following we assume that this range is much shorter than the radius of curvature everywhere on the surface; \ie, the interparticle distance $x_{\alpha\beta}\approx \big[ ( x_\alpha^{\ a}-x_\beta^{\ a} ) ( x_{\alpha a}-x_{\beta a} ) \big]^{1/2}$ wherever the kernel is nonzero. 

To understand the emergent behavior of this system, we apply
standard coarse graining techniques~\SI to Eq.~\eqref{eq:fp} to obtain hydrodynamic equations for the density $\rho \left( x,t\right) =\int duf\left( x,u,t\right) $ and the polarization $p^{a}\left( x,t\right) =\int duf\left( x,u,t\right) u^{a}$.
In dimensionless units such that $v_0=\Dr=1$, these equations read
\begin{align}
\label{eq:hydro_polar}
\begin{split}
& \partial_t \rho = - \nabla_a p^a
\\
& \partial_t p^a  =
- \frac12 \nabla^a \rho
+ \left( \kappa\rho - 1 - \frac{\kappa^2}{2} p^b p_b \right) p^a
\\ & \hspace{0.23\linewidth}
+ \frac{\kappa}{8} \left[ 5 p_b \nabla^a p^b
          - 5 p^a \nabla_b p^b - 3 p^b \nabla_b p^a \right]
\end{split}
\end{align}
where $\kappa=\int dr K(r)/2$.
They are formally identical to the equations reported in Ref.~\cite{Farrell2012} in the flat case, except for one important difference: the spatial derivatives become covariant derivatives.

In the absence of the aligning interaction ($\kappa=0$), the density evolves according to $\left( \partial_t^2 + \partial_t \right) \rho = \Delta \rho/2$,  where $\Delta$ is the curved-space Laplacian.
Therefore, barring constraints from the boundary, the system diffusively relaxes to homogeneity, with a diffusion constant $1/2$. 
The polarization obeys $\partial_t p^a = -\nabla^a\rho/2-p^a$ and relaxes to zero as the density becomes homogeneous.
In other words, in spite of the curvature-induced deflections, the steady state of a 2D ideal active gas is homogeneous and isotropic regardless of the shape of the surface on which it moves.

As in the flat case, this homogeneous isotropic state remains linearly stable in the presence of alignment as long as $\kappa\rho<1$. Above this threshold, the system spontaneously develops polar order. Unlike the flat case, however, there is no uniform polar solution, and finding a steady-state solution with polar order is highly non-trivial. Linear stability in turn cannot be studied without a base state to linearize about. 

To obtain a general steady-state solution with polar order to Eqs.~\eqref{eq:hydro_polar}, we now restrict ourselves to nearly flat surfaces and expand the density $\rho$ and polarization $p^a$ about the flat-surface homogeneous polar solution $(\R,\p^x=\sqrt{2\kappa\R-2}/\kappa,\p^y=0)$.
We assume a height profile $h(x,y)$ whose irregularities have typical height $H$ and width $W$ with $H/W\ll1$, thus a metric tensor $g_{ab}=\delta_{ab}+f_{ab}$ with $f_{ab}=(\partial_a h)(\partial_b h)\sim(H/W)^2\ll1$.
To leading order in $H/W$, the density deviation $\epsilon=\rho-\R\ll\R$ and the polarization deviation $\pi^a=p^a-\p^a\ll|\p|$ obey a set of third order, linear partial differential equations with constant coefficients and curvature dependent source terms.
Deep in the polar region ($W\kappa\p\gg1$), these equations reduce to a set of decoupled anisotropic Poisson equations, for which the Green's function $G$ is known. Thus $\epsilon$ and $\pi^a$ can be written explicitly in terms of the shape of the surface~\SI:
\begin{align}
&
\label{eq:solution_epsilon}
\epsilon (\vec{r}) = - \frac{\alpha \kappa \p^2}{2}
     ( \partial_y^2 F_{xx} + \partial_x^2 F_{yy} - 2 \partial_{xy} F_{xy} )
\\ &
\label{eq:solution_pix}
\pi^x (\vec{r})
= \frac{\partial\p}{\partial\R} \epsilon(\vec{r}) - \frac{\p}{2} f_{xx}(\vec{r})
\\ &
\label{eq:solution_piy}
\pi^y (\vec{r}) = - \frac{\p}{2} \left(
     - \partial_{xy} F_{xx} + \beta \partial_{xy} F_{yy} + 2 \partial_x^2 F_{xy} \right)
\end{align}
where
$F_{ab} (\vec{r}) = \int d\vec{r}' f_{ab} (\vec{r}') G(\vec{r}-\vec{r}')$,
$G(x,y) = \log\left( \beta x^2 + y^2 \right)/(4\pi\sqrt{\beta})$,
and $\beta=1/3$.
In the transition region, where polar order is present but weak ($0<W\kappa\p\ll1$), the same expressions hold except with $\beta=1$. 
Clearly, curvature affects both the density and the polarization, and this effect is nonlocal and linear in the metric deviation $f_{ab}$ (\ie quadratic in $H/W$).
Eq.~\eqref{eq:solution_pix} shows that $\pi^x$ only changes i) by following the changes of $\rho$ the same way the flat-space solution would and ii) to make up for the coordinates not being normalized any more.
As a result, the direction of polar order is modified, but its strength $|\langle\vec{u}\rangle|=|\vec{p}|/\rho$ remains unchanged (equal to $|\vec{\p}|/\R$), at least to leading order.
To assess the nonlocality, we compute the far field deviation caused by a localized surface deformation near the origin.
To leading order, the multipolar expansion of Eqs.~\eqref{eq:solution_epsilon}-\eqref{eq:solution_piy} reads
\newcommand{\favg}{\overline{f}}
\begin{align*}
\pi_x (\vec{r})
& = - \frac{\p}{4\pi\sqrt{\beta}} \
  \frac{ \left( \favg_{xx} - \beta \favg_{yy} \right) \left( \beta x^2 - y^2 \right)
         + 4 \beta \favg_{xy} x y
         }{ \left( \beta x^2 + y^2 \right)^2 }
\\
& = \frac{\epsilon (\vec{r})}{\kappa \p}
\\
\pi_y (\vec{r})
& = - \frac{\sqrt{\beta}\p}{2\pi} \
  \frac{ \left( \favg_{xx} - \beta \favg_{yy} \right) x y
         - \favg_{xy} \left( \beta x^2 - y^2 \right)
         }{ \left( \beta x^2 + y^2 \right)^2 }
\end{align*}
where $\favg_{ab}=\int d\vec{r}\, f_{ab}$.
The effect of a generic surface irregularity on the steady state density and polarization thus decays like $r^{-2}$; \ie, it is long ranged.


\sect{Strongly confined ideal active gas}
For our second example, let us consider a 3D self-propelled particle whose path is blocked by a hard frictionless wall. The wall simply cancels the normal component of the velocity of the particle. The equation of motion for the orientation of the particle in this case is $\frac{Du^a}{Dt}=-v_{0} u^\perp L^a_{\ b}u^b+\sqrt{\Dr}[\delta^a_{\ b}-u^a u_b/(1+u^\perp)]\xi^b$~\SI. Here $u^a u_a$ may take any value between $0$ (orientation normal to the wall) and $1$ (orientation tangent to the wall), $u^\perp=\sqrt{1-u_a u^a}$ is the component of the orientation normal to the wall, $L^a_{\ b}=\vec{e}^a\partial_b\vecu{n}$ is the shape tensor of the wall at the position of the particle, $\vecu{n}$ is the outward pointing normal to the wall, and $\xi$ is the same normalized Gaussian noise as before.

The effect of the wall is best explicated in the zero noise limit.
The eigenvalues of $L^a_{\ b}$ are the principal curvatures of the wall. The corresponding term in the equation of motion for $u$ describes the way the surface tilts in the particle's frame of reference.
Thus, the dynamics of $u^a$ is controlled by the sign of the principal curvatures. For convex walls (positive curvatures), $u^a$ relaxes to zero as the particle moves towards a location on the boundary where the normal is aligned with its orientation. If the wall is concave (negative curvatures), $u^a$ grows until $u^a u_a=1$ after which the particle points away from the wall and leaves it. Additionally, $L^a_{\ b}u^b$ in general produces an effective torque on the particle that aligns $u^a$ with the direction in which the curvature is smallest.

When the box is convex and the noise is weak (when the persistence length $v_0/\Dr$ is much larger than the box), we expect the orientation to relax toward the normal to the boundary much faster than it fluctuates ($u^a\ll 1$), and thus the particle never leaves the wall (which would require $u^a u_a=1$)~\cite{Fily2014a}. We show below that this approximation, which defines the \emph{strong confinement} regime, is self-consistent.
We first linearize the dynamics around $u^a=0$ to obtain $v^a=v_0 u^a$, $w^a=-v_0 L^a_{\ b} u^b$, and $\sigma^a_{\ b}=\sqrt{\Dr}\delta^a_{\ b}$. We then consider an ideal active gas, \ie, a collection of $N$ particles with negligible interparticle interactions. Expanding the \fp equation in moments of $u^a$ and neglecting third order moments then yields a closed system for the moments up to second order, which we solve to get~\SI
\begin{align}
\rho = \frac{N K}{4\pi}
, \
\left\langle u^a \right\rangle = 0
, \
\left\langle u^a u^b \right\rangle = \frac{\Dr}{v_0} R^{ab}
\label{eq:3d_results}
\end{align}
where
$K$ is the determinant of the shape tensor, \ie the Gaussian curvature, and
$R^a_{\ b}$ is the inverse of the shape tensor (its eigenvalues are the principal curvature radii).
In other words, density is proportional to the local Gaussian curvature, particles orient on average normal to the surface, and the fluctuations of the orientation around the normal in a given direction are controlled by the local radius of curvature along that direction.
When the active persistence length $v_0/\Dr$ is larger than both radii of curvature, $u^a$ remains small and particles stay on the boundary. This in turn proves the self-consistency of our approach.
Most importantly, the expression for $\rho$ can be used to predict the density of a 3D nonaligning ideal active gas on the boundary of its small confining box for arbitrary convex shapes (see the right panel of Fig.~\ref{fig:sketch} for an illustration). This is a key step toward designing active devices whose geometries are tuned to achieve specific functionalities. 
A more detailed treatment of this problem, including a comparison between our theory and the results of brownian dynamics simulations, will be published elsewhere.
Our approach should also be applicable to non-convex containers following the methods in Ref.~\cite{Fily2015}, however the transition from 2D to 3D will make the mathematics more involved. 


\sect{Conclusion}
In this paper we combine the tools of Riemannian geometry and nonequilibrium statistical mechanics to propose a statistical framework for self-propelled particles on arbitrary curved surfaces, which can be used to derive the statistical properties of the system from the microscopic dynamics and the geometry of the surface. We motivate the framework and demonstrate its use by determining the effect of surface curvature on two systems of interest: a 2D flocking model, and a 3D ideal active gas under strong convex confinement.

In addition to establishing the ability of the theory to solve diverse problems involving active particles on curved surfaces, these two examples highlight the variety of ways in which curvature can affect active systems. 
In our 2D model, the particles have no means to probe the geometry of the surface beyond what is required to make the theory self-consistent (\ie, independent of the choice of coordinates). This in turn severely limits their response to the surface geometry. At steady state, curvature only manifests itself when the system exhibits polar order. Additionally, the particles, not being aware the surface is embedded in 3D space, only respond to intrinsic curvature (\ie, the metric tensor and its derivatives).
Confined 3D particles, on the other hand, are sensitive to the angle their orientation makes with the tangent plane. By responding to the changes of this angle as they move along the surface, they can probe its curvature in specific directions -- in other words, they can sense both intrinsic and extrinsic curvature~\footnote{It so happens that the density in the zero noise limit is given by the Gaussian curvature, which is an intrinsic quantity. However fluctuations around this limit are controlled by the shape tensor, which is not.}. 
The importance of the intrinsic/extrinsic distinction to classifying active behaviors on curved surfaces comes from the fact that, like the symmetries (polar/nematic/isotropic) of activity and interparticle interactions~\cite{Marchetti2013}, it propagates from the individual to the collective dynamics. Our work provides a framework to study how it does so.


\begin{acknowledgments}
We thank Rastko Sknepnek and Silke Henkes for discussions. 
This work was supported by the NSF, DMR-1149266, and the Brandeis Center for Bioinspired Soft Materials, an NSF MRSEC, DMR-1420382.
Computational resources were provided by the NSF through XSEDE computing resources and the
Brandeis HPCC.
\end{acknowledgments}


\onecolumngrid
\section*{Appendix}
\newcommand{\refmain}[1]{#1}

\section{One-particle Fokker-Planck equation}

We start from the equations of motion Eqs.~\eqref{\refmain{eq:eom}} for $N$ particles on the curved surface:
\begin{align}
\forall a\in\{1,2\}, \ 
\forall \alpha\in[1,N], \ 
dx_\alpha^{\ a}/dt=v_\alpha^{\ a}, \
du^a/dt+\Gamma^a_{\ bc}(dx^b/dt)u^c=w_\alpha^{\ a}+\sigma_{\alpha\  b}^{\ a}\xi_\alpha^{\ a}
\label{ap:eq:eom}
\end{align}
where Greek indices are particle labels, latin indices are coordinate labels, 
$\xi$ is a Gaussian noise with zero mean and correlations $\left\langle \xi_{\alpha}^{\ a}(t) \xi_{\beta b}(t') \right\rangle =2\delta_{\alpha\beta} \delta_{\ b}^{a} \delta(t-t') $, $\Gamma^a_{\ bc}$ are the Christoffel symbols of the second kind, and $v$,
$w$, $\sigma $ are arbitrary functions of the positions and orientations
of the particles. 
The usual (noncovariant) one-particle phase space density
\newcommand{\f}{\tilde{f}}
\begin{align}
\f(x,u,t) 
= \Big\langle \sum_\alpha
\prod_a \delta\left[ x^a - X_\alpha^{\ a}(t) \right] 
\delta\left[ u^a - U_\alpha^{\ a}(t) \right]
\Big\rangle
\end{align}
obeys the usual one-particle Fokker-Planck equation
\begin{align}
\partial_t \f = - \partial_{x^a} \left( \mf{v}^a \f \right)
    - \partial_{u^a} \left[ \left( - \Gamma^a_{\ bc} \mf{v}^b u^c + \mf{w}^a \right) \f \right]
    + \partial_{u^a} \left[ \mf{\sigma}^{ac} \partial_{u^b} \left( \mf{\sigma}_c^{\ b} \f \right) \right]
\end{align}
where overbars denote averages over the trajectories $\left\{\left( X_{\alpha }\left( t\right)
,U_{\alpha }\left( t\right) \right)\right\}_\alpha$ of all the other particles. 
The equation is closed by a mean-field approximation wherein the multi-particle distribution functions appearing in these averages are written as products of one-particle distribution functions, \ie, correlations between particles are neglected:
\begin{align}
\forall z \in \{v,w,\sigma\}, \ 
\mf{z}^a (x,u,t) 
\approx 
\int \left(\prod_{\beta,b} dx_\beta^b du_\beta^b\right)
\left( \prod_\beta \f(x_\beta,u_\beta,t) \right)
\sum_\alpha z_\alpha^{\ a} (\{x_\beta\},\{u_\beta\}) 
   \prod_b \delta(x^b-x_\alpha^{\ b}) \delta(u^b-u_\alpha^{\ b})
\end{align}
The scalar (\ie coordinate independent) density $f=\f/g$ where $g(x)$ is the determinant of the metric thus obeys
\begin{align}
\partial_t f = 
    - \partial_{x^b} \left( \mf{v}^a f \right) - \Gamma^a_{\ bc} \mf{v}^c f + \Gamma^d_{\ bc} u^c \partial_{u^d} \left( \mf{v}^a f \right)
    - \partial_{u^a} \left( \mf{w}^a f \right)
    + \partial_{u^a} \left[ \mf{\sigma}^{ac} \partial_{u^b} \left( \mf{\sigma}_c^{\ b} f \right) \right]
\label{ap:eq:fp}
\end{align}
where we have used the relationship $\partial_{x^a} g = g g^{bc} \partial_{x^a} g_{bc} = 2 g \Gamma^b_{\ ba}$ and the fact that $g_{ab}$ doesn't depend on $u$. 
The first three terms on the right-hand side form minus the phase space covariant derivative $\covps\left( \mf{v}^a f \right)$ defined by Eq.~\eqref{\refmain{eq:covdev}}.


\section{Moment expansion}
\label{ap:moments}

\newcommand{\uprod}{\prod_{m=1}^n u^{a_m}}

The hydrodynamic equations are obtained from the first few moments in the orientation $u$ of Eq.~\eqref{ap:eq:fp}, the first of which is the density field $\rho$:
%
%
\begin{align}
\rho(x,t) \equiv \int du\, f(x,u,t)
, \quad 
\left\langle \prod_{m=1}^n u^{a_m} \right\rangle(x,t)  \equiv
\frac{1}{\rho(x,t)}\int du\, f(x,u,t) \prod_{m=1}^n u^{a_m}
\end{align}
where $du=\sqrt{g(x)}\prod_a du^a$ is the covariant orientational volume element.
Multiplying the Fokker-Planck equation by $\prod_{m=1}^n u^{a_m}$, integrating over $u$, and integrating by part as needed to eliminate derivatives of $f$ leads to
%
\begin{multline}
\forall n, \ 
\partial_t \left( \rho \left\langle \uprod \right\rangle \right)
   = - \nabla_b \left( \rho \left\langle \mf{v}^b \uprod \right\rangle \right)
     + \sum_{p=1}^n \rho \left\langle \mf{w}^{a_p} \prod_{\substack{m=1\\m\neq p}}^n u^{a_m} \right\rangle
     \\
     + \sum_{p=1}^n \rho \left\langle \mf{\sigma}^b_{\ c} \left(\partial_{v^b} \mf{\sigma}^{c a_p}\right) \prod_{\substack{m=1\\m\neq p}}^n u^{a_m} \right\rangle
     + \sum_{\substack{p,q=1\\p\neq q}}^n \rho \left\langle \mf{\sigma}^{a_p}_{\ \ c} \mf{\sigma}^{c{a_q}} 
          \prod_{\substack{m=1\\m\neq p,q}}^n u^{a_m} \right\rangle
\label{ap:eq:moment_dynamics}
\end{multline}
%
%
where $\nabla$ is the usual covariant derivative.
Since the moments are by construction contravariant tensors of order $n$, Eq.~\eqref{ap:eq:moment_dynamics} is explicitly covariant.
It is not in general a closed set of equations for the moments, however it is possible to make it one by expanding $\mf{v}^a$, $\mf{w}^a$, and $\mf{\sigma}^{ab}$ in powers of $u^a$.
%


\section{Flying XY model}

\subsection{Coarse graining}

Apart from curvature, our equations of motion are identical to those from Ref.~\cite{Farrell2012}, and we use approximations that are equivalent to theirs. 
We, however, work with the coordinates of $u$ in the surface coordinate system rather than its angle with a reference axis as the latter would be tricky to define on a curved surface.
Let $p^a(x,t)=\rho\left\langle u^a \right\rangle$ and $q^{ab}(x,t)=\rho\left(\left\langle u^a u^b\right\rangle - \frac12 g^{ab}\right)$ be the polarization field and nematic tensor field, respectively.
To leading order in the short range of the aligning interaction, the mean-field torque is
\begin{align}
\mf{w^a} = \int dx' du' (\delta^a_{\ b}-u^a u_b) u'^b K(|x-x'|) \f(x,u,t) \f(x',u',t) 
   \approx K_0 (\delta^a_{\ b}-u^a u_b) \f(x,u,t) p^b(x,t)
\end{align}
where $K_0=\int dx K(x)$ and $dx=\sqrt{g(x)}\prod_a dx^a$ is the covariant volume element. 
Substituting this expression as well as $v^a=v_0 u^a$ and 
$\sigma^a_{\ b}=\sqrt{\Dr}( \delta^a_{\ b} -u^a u_b)$ in Eqs.~\eqref{ap:eq:moment_dynamics}
and using the covariant form of the standard 2D third order moment closure 
$\rho\left\langle u^a u^b u^c\right\rangle=(g^{ab}p^c+g^{ac}p^b+g^{bc}p^a)/4$~\cite{Ahmadi2006} yields
\begin{align}
& \partial_t \rho + v_0 \nabla_a p^a = 0 \\
& \partial_t p^a + \frac{v_0}{2} \nabla^a \rho + v_0 \nabla_b q^{ab} 
  = -\Dr p^a + K_0 \left( \frac12 \rho p^a - q^{ab} p_b \right) \\
& \partial_t q^{ab} + \frac{v_0}{4} \left( \nabla^a p^b + \nabla^b p^a - g^{ab} \nabla_c p^c \right)
  = -4\Dr q^{ab} + K_0 \left( p^a p^b - g^{ab} p^c p_c \right)
\end{align}
The hydrodynamic equations are then obtained by assuming that $q^{ab}$ is a fast variable ($\partial_t q^{ab}\approx0$) and that the hydrodynamic fields vary slowly (neglecting spatial derivatives of order two and higher):
\begin{align}
\label{ap:eq:hydro_polar}
\begin{split}
& \partial_t \rho = - \nabla_a p^a
\\
& \partial_t p^a  =
- \frac{v_0}{2} \nabla^a \rho
+ \left( \frac{K_0}{2}\rho - \Dr - \frac{K_0^2}{2\Dr} p^b p_b \right) p^a
+ \frac{K_0 v_0}{16\Dr} \left[ 5 p_b \nabla^a p^b
          - 5 p^a \nabla_b p^b - 3 p^b \nabla_b p^a \right]
\end{split}
\end{align}
Finally we rescale times and lengths using the reorientation time $\Dr^{-1}$ and the persistence length $v_0/\Dr$ to get the dimensionless hydrodynamic equations Eqs.~\eqref{\refmain{eq:hydro_polar}}.


\subsection{Polar steady-state on a weakly curved surface}

Here we assume the surface is described by a height profile $h(x,y)$ whose irregularities have typical height $H$ and width $W$ with $H/W\ll1$.
Then the metric tensor is $g_{ab}=\delta_{ab}+f_{ab}$ with $f_{ab}=(\partial_a h)(\partial_b h)\sim(H/W)^2\ll1$ and the Christoffel symbols are $\Gamma_{abc}\sim\partial_a f_{bc}\sim H^2/W^3$. 
Let $\epsilon=\rho-\R\ll\R$ and $\pi^a=p^a-\p^a\ll|\p|$ be the density and polarization deviations from the flat-space mean-field steady-state. 
We assume we are at steady-state in the polar phase ($\kappa\R>1$) and that the flat-space base state is polarized along the $x$ axis: $\p^x=\sqrt{2\kappa\R-2}/\kappa$, $\p^y=0$.
We first note that the density conservation equation 
$0=\nabla_a p^a\approx\nabla_a\pi^a+\Gamma^b_{\ ba}\p^b$ 
implies $\pi^a\sim(H/W)^2\p$, then expand the dimensionless hydrodynamic equations Eqs.~\eqref{\refmain{eq:hydro_polar}} to leading order in $H/W$:
\begin{align}
\label{eq:linearized_polar_1}
& 2 \partial_x \pi_x + 2 \partial_y \pi_y
= - \p \partial_x \left( f_{xx} + f_{yy} \right)
\\
\label{eq:linearized_polar_2}
& 4 \left( 2 \kappa \p - \partial_x \right) \epsilon 
- 2 \kappa \p \left( 4 \kappa \p - \partial_x \right) \pi_x
= \kappa \p^2 \left( 4 \kappa \p - \partial_x \right) f_{xx}
\\
\label{eq:linearized_polar_3}
&  - 4 \partial_y \epsilon 
+ \kappa \p \left( 5 \partial_y \pi_x - 3 \partial_x \pi_y \right)
= \kappa \p^2 \left( 3 \partial_x f_{xy} - 4 \partial_y f_{xx} \right)
\end{align}
where all the Christoffel symbols have been expressed in terms of the metric using 
$\Gamma_{abc}=\frac12(\partial_c g_{ab}+\partial_b g_{ac}-\partial_a g_{bc})$. 
We further simplify \eqref{eq:linearized_polar_2} by assuming either strong polar order ($\kappa\p W\gg 1$ thus $\kappa \p \gg \partial_x$) or weak polar order ($\kappa\p W\ll 1$ thus $\kappa \p \ll \partial_x$).
In both cases Eqs.~\eqref{eq:linearized_polar_1}-\eqref{eq:linearized_polar_3} reduce to a set of anisotropic Poisson equations:
\newcommand{\myOp}{\left(\partial_x^2 + \beta \partial_y^2\right)}
\begin{align}
\label{eq:decoupled_polar_epsilon}
& \myOp \epsilon = - \frac{\kappa \p^2}{2}
  \left( \partial_y^2 f_{xx} + \partial_x^2 f_{yy} - 2 \partial_{xy} f_{xy} \right)
\\
\label{eq:decoupled_polar_pix}
& \myOp \pi^x = - \frac{\p}{2} 
  \Big( \left[ \partial_x^2 + (1+\beta) \partial_y^2 \right] f_{xx} 
     + \partial_x^2 f_{yy} - 2 \partial_{xy} f_{xy} \Big)
\\
\label{eq:decoupled_polar_piy}
& \myOp \pi^y = - \frac{\p}{2} 
  \Big( - \partial_{xy} f_{xx} + \beta \partial_{xy} f_{yy} + 2 \partial_x^2 f_{xy} \Big)
\end{align}
where $\beta=1$ if $\kappa\p W\ll 1$ and $\beta=1/3$ if $\kappa\p W\gg 1$. 
The solution can then be given an explicit expression in terms of the source terms, hence the metric, using the well known Green's function for $\myOp$ in 2D unbounded space $G(x,y)=\log\left( \beta x^2 + y^2 \right)/(4\pi\sqrt{\beta})$, which satisfies $\myOp G = \delta(x)\delta(y)$:
\begin{align}
u (x,y) = \int dx'dy' U(x-x',y-y') \frac{\log\left( \beta x'^2 + y'^2 \right)}{4\pi\sqrt{\beta}} 
\end{align}
where $u$ is either $\epsilon$ or $\pi^a$ and $U$ is the corresponding source. 
Finally, taking the derivatives appearing in the source terms out of the integral yields Eqs.~\eqref{\refmain{eq:solution_epsilon}}-\eqref{\refmain{eq:solution_piy}}.


\section{Confined 3D active particles}

\subsection{Equations of motion on the boundary}
\label{ap:eom}

We consider a single 3D active particle with position $\vec{r}$ and unit orientation vector $\vecu{u}$ moving along a wall that cancels the normal component of its velocity whenever it points toward the wall. The equations of motion are
\begin{align}
\frac{d\vec{r}}{dt} = v_0 \left( 1 - \vecu{n} \otimes \vecu{n} \right) \vecu{u}
, \
\frac{d\vecu{u}}{dt} =
\vecu{u} \times \left( \vecg{\eta}\times\vecu{u} \right)
\label{ap:eq:eom_3d}
\end{align}
where $v_0$ is the self-propulsion speed, $\vecu{n}$ is the normal to the wall at $\vec{r}$, $\vecg{\eta}$ is a white Gaussian noise with zero mean and correlations $\langle\eta_i(t)\eta_j(t')\rangle=2\Dr\delta_{ij}\delta(t-t')$, and latin letters from the middle of the alphabet ($i$, $j$,...) label 3D cartesian coordinates (whereas latin letters from the beginning of the alphabet label coordinates on the surface).
We now project Eqs.~\eqref{ap:eq:eom_3d} onto the local basis $(\vec{e}_1,\vec{e}_2)$ of the plane tangent to the surface:
%
%
\begin{align}
& \vec{e}^a\cdot\dfrac{d\vec{x}}{dt}=\dfrac{dx^a}{dt}
\\
& \vec{e}^a\cdot v_0\vecu{u} = v_0 u^a
\\
& \vec{e}^a\cdot\frac{d\vecu{u}}{dt}
  = \vec{e}^a \cdot \left( \frac{du^b}{dt} \vec{e}_b + u^b \frac{d\vec{e}_b}{dt} + \frac{du^\perp}{dt} \vecug{n} + u^\perp \frac{d\vecu{n}}{dt}  \right)
  = \frac{du^a}{dt} + v_0 \Gamma^a_{\ bc} u^b u^c + v_0 u^\perp L^a_{\ b} u^b 
\label{ap:eq:nu}
\\ & \nonumber \\
& \vec{e}^a \cdot \big[ \vecu{u} \times \left( \vecg{\eta} \times \vecu{u} \right) \big]
  = \left[ \left(\delta^a_{\ b}-u^a u_b\right) \vec{e}^b - u^\perp u^a \vecug{n} \right]\cdot\vecg{\eta}
  \equiv \Xi^a
\label{ap:eq:noise}
\end{align}
where $u^\perp=\vecu{u}\cdot\vecu{n}=\sqrt{1-u_a u^a}$ 
and we have used
$\vec{e}^a\cdot\partial_b\vec{e}_c = \Gamma^a_{\ bc}$ and 
$\vec{e}^a\cdot\partial_b\vecu{n} = -\vecu{n}\cdot\partial_b\vec{e}^a = L^a_{\ b}$.
The noise $\Xi^a(t)$ defined by Eq.~\eqref{ap:eq:noise} is Gaussian with zero mean and correlations
\begin{align}
\left\langle \Xi^a(t)\, \Xi^b(t') \right\rangle
  = 2\Dr \left( g^{ab} - u^a u^b \right) \delta(t-t')
\label{ap:eq:variance}
\end{align}
which we rewrite as $\Xi^a = \sigma^a_{\ b} \xi^b$ where $\xi$ is a Gaussian noise with zero mean and correlations 
$\left\langle \xi^a(t)\, \xi^b(t') \right\rangle = 2g^{ab}\delta(t-t')$ and 
\begin{align}
\sigma^a_{\ b} = \sqrt{\Dr} \left( \delta^a_{\ b} - \frac{u^a u_b}{1+u^\perp} \right)
\label{ap:eq:sigma}
\end{align}
Therefore
\begin{align}
\frac{Du^a}{Dt} = \frac{du^a}{dt} + \Gamma^a_{\ bc}\frac{dx^b}{dt}u^c = 
- v_0 u^\perp L^a_{\ b} u^b 
+ \sqrt{\Dr} \left( \delta^a_{\ b} - \frac{u^a u_b}{1+u^\perp} \right) \xi^b
\label{ap:eq:eom_confA}
\end{align}
%


\subsection{Steady-state in the strong confinement regime}
\label{ap:confined}

In the strong confinement regime $u^a u_a\ll1$ and we linearize Eq.~\eqref{ap:eq:eom_confA} in $u$ (except for the intrinsic time derivative whose linearization would break covariance):
\begin{align}
\frac{dx^a}{dt} = v_0 u^a 
, \ 
\frac{Du^a}{Dt} = 
- v_0 L^a_{\ b} u^b 
+ \sqrt{\Dr} \xi^a
\label{ap:eq:eom_confA_linearized}
\end{align}
To obtain the steady-state density we substitute the corresponding expression for $v$, $w$, $\sigma$ in Eqs.~\eqref{ap:eq:moment_dynamics}, set $\partial_t=0$, and neglect third order moments ($\left\langle u^a u^b u^c\right\rangle\approx0$)~\cite{Fily2014a}. 
The first three equations then read
\begin{align}
& 0 = -\nabla_a \left( \rho \left\langle u^a \right\rangle \right)
, \quad
0 = -\nabla_b \left( \rho \left\langle u^a u^b \right\rangle \right) + v_0 L^a_{\ b} \rho \left\langle u^b \right\rangle
, \quad
0 = L^a_{\ c} \left\langle u^c u^b \right\rangle + L^b_{\ c} \left\langle u^c u^a \right\rangle + 2 \frac{\Dr}{v_0} g^{ab}
\label{ap:eq:moments_strong3}
\end{align}
Since $g^{ab}$, $L^{ab}$, $\left\langle u^a u^b \right\rangle$ are all symmetric and $g^{ab}$ and $L^{ab}$ have only positive eigenvalues and commute with each other, $L^{ab}$ also commutes with $\left\langle u^a u^b \right\rangle$ and the third equation reduces to 
\begin{align}
\left\langle u^a u^b \right\rangle = \frac{\Dr}{v_0} R^{ab}
\label{ap:eq:variance_strong3}
\end{align}
where $R^a_{\ b}$ is the inverse of the shape tensor $L^a_{\ b}$. 
Next we look for a fluxless ($\left\langle u^a \right\rangle=0$) solution to Eqs.~\eqref{ap:eq:moments_strong3}.
Substituting Eq.~\eqref{ap:eq:variance_strong3} into the middle equation and using the fluxless assumption leads to 
\begin{align}
\nabla_b \left( R^{ab} \rho \right) = 0
\label{eq:density_eq}
\end{align}
\newcommand{\xo}{\overline{x}}%
We now pick a point $\xo$ on the boundary and go to its local normal coordinates, in which $\nabla_a = \partial_a$ and $R^{ab}$ is diagonal. In the vicinity of $\xo$ the radius of curvature tensor can be written as
\begin{align*}
\tens{R}(x)= \tens{P}^{-1} \cdot \left(\begin{array}{cc} R_1(x) & 0 \\ 0 & R_2(x) \end{array}\right) \cdot \tens{P}
, \quad 
\tens{P}=\left(\begin{array}{cc} \cos\phi(x) & \sin\phi(x) \\ -\sin\phi(x) & \cos\phi(x) \end{array}\right)
\end{align*}
where $R_{1,2}$ are the principal curvature radii and $\phi$ is the angle between the curvature eigenbasis at $x$ and that at $\xo$, which by construction verifies $\phi(\xo)=0$.
With these notations, we have 
\begin{align}
\vecg{\nabla}\cdot \tens{R} (\xo) =
  \left(\begin{array}{c}
     \partial_1 R_1 + (R_1-R_2)\partial_2\phi \\
     \partial_2 R_2 + (R_1-R_2)\partial_1\phi
  \end{array}\right)
\label{ap:eq:curv_div1}
\end{align}
We use the Mainardi-Codazzi equations, which follow from $\partial_{bc}\vec{e}_a=\partial_{cb}\vec{e}_a$, to eliminate $\phi$: 
\begin{align}
\nabla_c L_{ab}=\nabla_b L_{ac}
\implies
  \left\{
  \begin{array}{c}
     \left(R_1-R_2\right) \partial_2\phi =  \dfrac{R_1}{R_2} \partial_1 R_2
     \\
     \left(R_1-R_2\right) \partial_1\phi =  \dfrac{R_2}{R_1} \partial_2 R_1
  \end{array}
  \right.
\end{align}
Substituting in Eq.~\eqref{ap:eq:curv_div1}, we find
\begin{align}
\vecg{\nabla}\cdot \tens{R} (\xo)
= \left(\begin{array}{c}
     R_1\, \partial_1 \left[ \ln\left( R_1 R_2 \right) \right] \\
     R_2\, \partial_2 \left[ \ln\left( R_1 R_2 \right) \right]
  \end{array}\right)
= - \tens{R} \cdot \vecg{\nabla} \left( \ln K \right)
\end{align}
where $K\equiv 1/(R_1 R_2) = \det \tens{L}$ is the Gaussian curvature.
The final result is explicitly covariant and holds in any coordinate system.
Applying it to Eq.~\eqref{eq:density_eq} yields
\begin{align}
0 = \rho \left( \nabla_j R^{ij} \right) + R^{ij} \left( \nabla_j \rho \right)
  = \rho R^{ij} \nabla_j \ln \left( \frac{\rho}{K} \right)
\end{align}
Integrating and using the normalization constraint $\oiint \rho\, dS = N$
where $N$ is the number of (independent) particles on the boundary and the Gauss-Bonnet theorem
$\oiint K\, dS = 4\pi$ for a closed surface with no handles leads to 
\begin{align}
\rho = \frac{N K}{4\pi}
\end{align}
Plugging this result back into Eqs.~\eqref{ap:eq:moments_strong3} proves the existence of a fluxless steady-state that verifies Eqs.~\eqref{\refmain{eq:3d_results}}.


\bibliography{curved_active}

\end{document}